\documentclass[reprint,superscriptaddress,nofootinbib,amsmath,amssymb,prd,floatfix]{revtex4-1}
\usepackage{graphicx}
\usepackage{bm}
\newcommand*{\FigPath}{./figures}%
\usepackage[colorlinks=true,linkcolor=blue,bookmarksopen,bookmarksnumbered]{hyperref}
\def\bea#1\eea{\begin{align}#1\end{align}} 

\newcommand{\bef}{\begin{figure}[htb]\centering}
\newcommand{\eef}{\end{figure}}
\let\vec\bm
\def\L{\Lambda}

\newcommand{\fref}[1]{Fig.~\ref{f.#1}}

\newcommand{\sref}[1]{Sec.~\ref{s.#1}}

\newcommand{\tref}[1]{Table~\ref{t.#1}}

\begin{document}
	
\title{Extracting the Transverse Momentum Dependent Polarizing Fragmentation Functions}
	
\author{Daniel Callos}
\email{dcallos@g.ucla.edu}
\affiliation{Department of Physics and Astronomy, University of California, Los Angeles, California 90095, USA}

\author{Zhong-Bo Kang}
\email{zkang@physics.ucla.edu}
\affiliation{Department of Physics and Astronomy, University of California, Los Angeles, California 90095, USA}
\affiliation{Mani L. Bhaumik Institute for Theoretical Physics, University of California, Los Angeles, California 90095, USA}
\affiliation{Center for Frontiers in Nuclear Science, Stony Brook University, Stony Brook, New York 11794, USA}

\author{John Terry}
\email{johndterry@physics.ucla.edu}
\affiliation{Department of Physics and Astronomy, University of California, Los Angeles, California 90095, USA}
\affiliation{Mani L. Bhaumik Institute for Theoretical Physics, University of California, Los Angeles, California 90095, USA}

\begin{abstract} 
We demonstrate that spontaneous transverse polarization of Lambda baryon ($\Lambda$) production in $e^+e^-$ annihilation can be described using the transverse momentum dependent polarizing fragmentation functions (TMD PFFs). Using a simple Gaussian model, we perform an extraction of the TMD PFFs by fitting the BELLE collaboration's recent measurement of the $\Lambda$ transverse polarization in back-to-back $\Lambda+h$ production in $e^+ e^-$ collisions, $e^{-} + e^{+} \rightarrow \Lambda^{\uparrow}+h+X$. We find that this simple model accurately describes the experimental data for $\Lambda$ production associated with pions and kaons, and we are able to determine TMD PFFs for different quark flavors. We use these newly extracted TMD PFFs to make predictions for the transverse polarization of $\Lambda$ produced in semi-inclusive deep inelastic scattering at a future electron-ion collider, and find that such a polarization is around $10\%$ and should be measurable.
\end{abstract}
\maketitle

\section{Introduction}\label{s.Introduction}

The first observation of large transverse single-spin asymmetries (SSAs) in Lambda baryon ($\Lambda$) production was made more than forty years ago \cite{Bunce:1976yb}.  As the consensus of the time was that such QCD spin effects at colliders should be small~\cite{Kane:1978nd}, this experimental discovery came as a surprise to the scientific community. These discoveries demonstrated that a detailed description of transverse spin physics was essential for a high-precision understanding of collider data. 

Tremendous progress has been made in the past decades, with the help of QCD factorization theorems
\cite{Collins:1989gx,Qiu:1991pp,Qiu:1991wg,Collins:1981uk,Collins:1984kg}. For example, a recent phenomenological
analysis presented in~\cite{Cammarota:2020qcw} demonstrates that single transverse-spin asymmetries for light hadrons, such as pions in high-energy collisions, have a common origin. Namely, they are due to the intrinsic quantum-mechanical interference from multi-parton states in the parent proton and/or in the fragmenting hadron. 

One of these quantum-mechanical interferences is encoded in the Sivers function~\cite{Sivers:1989cc}, which describes the distribution of unpolarized quarks inside a transversely polarized proton, through a correlation between the parton's transverse momentum with respect to the proton direction and the transverse spin vector of the proton. The exact same interference can arise in the hadronization process, giving rise to the so-called \textit{polarizing fragmentation functions} (PFFs). The PFFs describe an unpolarized quark that fragments into a transversely polarized spin-1/2 hadron, such as a $\Lambda$ baryon. In this case, the PFFs encode the correlation between the hadron's transverse momentum with respect to the fragmenting quark and the transverse spin of the $\Lambda$ particle. 

While the origin of $\Lambda$ polarization has been an active field of study for the past forty years, data have been available mainly from single inclusive $\Lambda$ production in proton-proton collisions, $p p \rightarrow \Lambda +X$. For such a process with a single hard scale -- in this case the transverse momentum of $\Lambda$ -- one can establish a QCD collinear factorization formalism at high-twist~\cite{Koike:2017fxr,Gamberg:2018fwy}. For processes with more than one hard scale, a transverse momentum dependent (TMD) factorization is to be used~\cite{Collins:2011zzd}. This would be the case for studying the polarization of $\Lambda$ production in semi-inclusive deep inelastic scattering (SIDIS), as well as back-to-back $\Lambda+h$ production in $e^+e^-$ collisions. In both processes, the transverse momentum dependent polarizing fragmentation functions (TMD PFFs) $D_{1T}^\perp$ could be studied. However experimental data has not been available for either of these processes until recently.

The BELLE collaboration has recently measured the transverse polarization of the $\Lambda$ in $e^+e^-$ annihilation~\cite{Guan:2018ckx}. They have measured such a polarization of both $\L$ and $\bar{\L}$ in single $\Lambda$ production (with respect to the thrust axis), $e^{-} + e^{+} \rightarrow \Lambda/\bar{\L}+X$, as well as the back-to-back $\Lambda+h$ production, $e^{-} + e^{+} \rightarrow \Lambda/\bar{\L}+h+X$. While a well-established TMD factorization formalism exists to treat back-to-back $\Lambda+h$ production~\cite{Collins:1981uk,Collins:2011zzd}, single $\Lambda$ production with respect to the thrust axis could involve a more complicated factorization structure~\cite{Jain:2011iu}, if the thrust variable is also measured. Nevertheless, there is an attempt at factorization within the standard TMD formalism~\cite{Anselmino:2019cqd} for single $\Lambda$ production. In this paper, we will focus on back-to-back $\Lambda+h$ production because of this extra complication. 

These BELLE data allow for the extraction of the TMD PFFs. This extraction is a major goal of the TMD community, as it represents one of eight leading-twist TMDs for the TMD FFs, and thus provides three-dimensional imaging of hadrons in association with the fragmentation process. Furthermore, a high-precision description of the TMD PFFs is vital to our understanding of correlations between final-state hadron spin and intrinsic transverse momenta of the elementary constituents. The understanding of these spin-transverse momentum correlations gives rise to interesting phenomenological differences between TMD FFs and the TMD parton distribution functions (PDFs). 

For instance, the Sivers functions, TMD PDFs analogous to the TMD PFFs, exhibit so-called \textit{modified universality} -- a sign change -- between the SIDIS and Drell-Yan processes~\cite{Collins:2002kn,Boer:2003cm,Kang:2009bp}. While the TMD PFF is T-odd just like the Sivers function, this TMD does not exhibit modified universality between SIDIS and $e^{-} + e^{+} \rightarrow \Lambda+h+X$; rather, the TMD PFF should be universal with respect to these two processes~\cite{Metz:2002iz,Collins:2004nx,Meissner:2008yf,Boer:2010ya}. In fact, Ref.~\cite{Boer:2010ya} has precisely suggested studies of both back-to-back $\Lambda+h$ production and SIDIS to test the universality of the TMD PFFs. In this paper, we provide a prediction for the transverse polarization in SIDIS, which can be used for the first experimental confirmation of the universality of the TMD PFFs.

Within the TMD factorization formalism, we perform an extraction of the TMD PFFs, from the recent $\Lambda/\bar{\Lambda}$ polarization measurements recorded at BELLE~\cite{Guan:2018ckx}. We study in detail the implications of the TMD PFFs for different quark flavors, and provide predictions for the $\Lambda/\bar{\Lambda}$ polarization in SIDIS. We organize our work as follows. In \sref{Formalism} we provide the relevant formalism and detail the calculation of the $\Lambda$ transverse polarization observable $P_{\perp}^{\Lambda}$. In \sref{Fit Results} we give the parametrization of our TMD PFFs and discuss the fit procedure, fit results, and our predictions for SIDIS. We conclude the paper in \sref{Conclusions}.

\section{Formalism}
\label{s.Formalism}
In this section, we provide the QCD formalism for describing $\Lambda$ polarization. 
We consider back-to-back production of a $\Lambda$ baryon and a light hadron $h$ in the final state, 
\bea
e^-(\ell)+e^+(\ell^\prime) \to \gamma^{*}(q) \to  h(P_h)+\Lambda(P_{\Lambda}, \vec{S}_{\perp})+X,
\eea
where $q=\ell + \ell^\prime$ is the momentum of the intermediate virtual photon with $q^2\equiv Q^2$, and we denote the momentum of the outgoing light hadron and the $\Lambda$ by $P_h$ and $P_{\Lambda}$, respectively. We further define
\bea
z_\Lambda = 2P_\Lambda\cdot q/Q^2, 
\qquad
z_h = 2P_h\cdot q/Q^2. 
\eea
Following~\cite{Boer:1997mf}, we choose a leptonic center-of-mass frame where the light hadron $P_h$ has no transverse momentum. The leptons and the light hadron form the so-called \textit{leptonic plane}. 
The angle between $P_h$ and $(\ell,~\ell')$ is given by $\theta$, as illustrated in Fig.~\ref{f.frame}. 
%
\bef
\includegraphics[width = 0.9\linewidth]{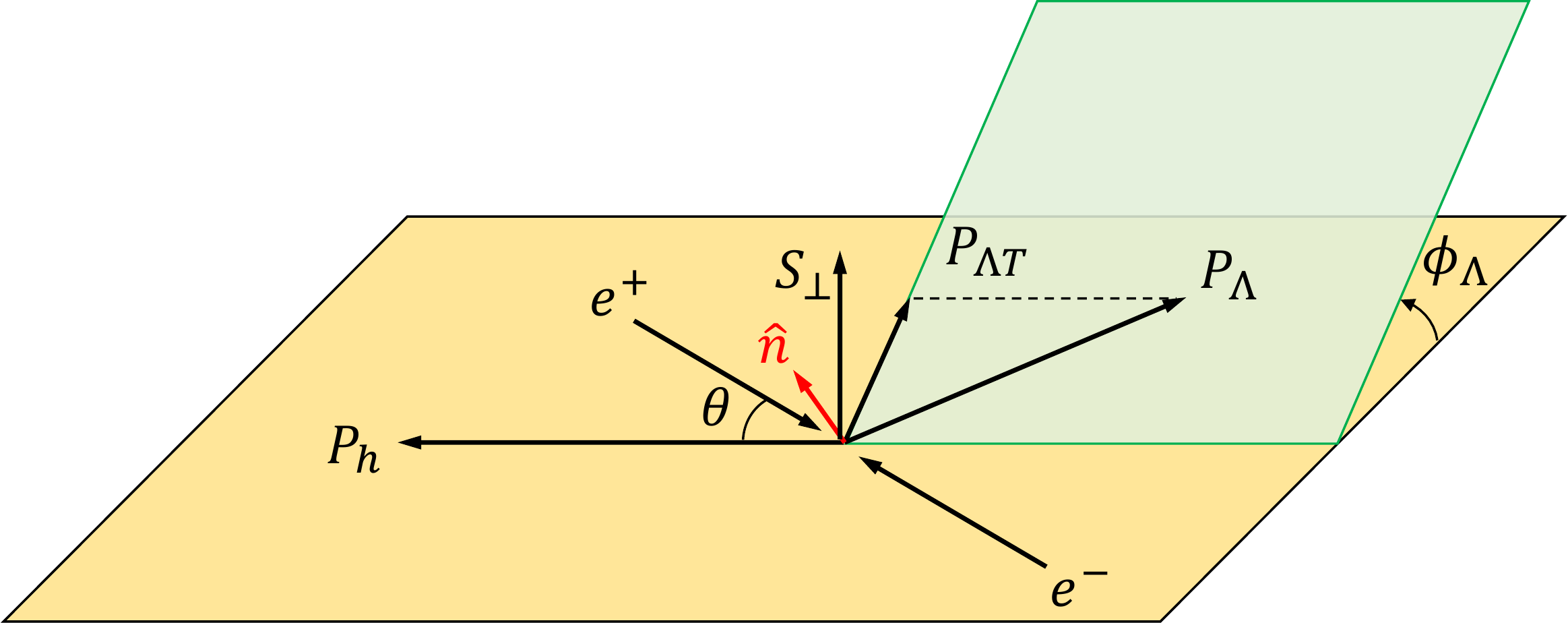}
\caption{Kinematics of the leptonic center-of-mass frame for back-to-back two-hadron production in $e^-e^+$ annihilation, $e^-+e^+\to h(P_h)+\Lambda(P_\Lambda)+X$.}
\label{f.frame}	
\eef
On the other hand, $P_h$ and $P_\Lambda$ span the so-called \textit{hadronic plane}. 
In this frame, the $\Lambda$ particle has transverse momentum $\vec{P}_{\Lambda T}$, at an azimuthal angle $\phi_\Lambda$ with respect to the leptonic plane. 
We have
\bea
\vec{P}_{\Lambda T} = -z_\Lambda \vec{q}_\perp,
\eea
where $\vec{q}_\perp$ is related to the ``transverse'' component of the virtual photon momentum, defined as
\bea
q_t^\mu = q^\mu - \frac{P_h\cdot q}{P_h\cdot P_\Lambda} P_\Lambda^\mu - \frac{P_\Lambda\cdot q}{P_\Lambda \cdot P_h} P_h^\mu,
\eea
with $q_\perp^2 = -q_t^\mu q_{t\mu}$.

We start with the QCD factorization formalism for the unpolarized differential cross section~\cite{Collins:1981uk,Boer:2010ya}
\bea
\frac{d\sigma}{d\mathcal{PS} d^2 \vec{q}_\perp} =& \sigma_0 H(Q) z_\Lambda^2z_h^2 \sum_q e_q^2 
\int d^2 \vec{k}_{h \perp} d^2 \vec{k}_{\Lambda \perp} d^2 \vec{\lambda}_{\perp} 
\nonumber \\
&\times\delta^{(2)}(\vec{k}_{\Lambda \perp}+\vec{k}_{h \perp}+\vec{\lambda}_{\perp}-\vec{q}_{\perp}) S(\vec{\lambda}_{\perp})
\nonumber\\
&\times D_{\Lambda/q}(z_\Lambda, p_{\Lambda\perp}^2) D_{h/\bar{q}}(z_h, p_{h\perp}^2)\,,
\label{e.factorization.0}
\eea
where $d\mathcal{PS} =dz_\Lambda\, dz_h \,d(\cos\theta)$ and $\sigma_0$ is given by
\bea
\sigma_0 = \frac{N_c\pi \alpha_{\rm em}^2}{2Q^2}\left(1+\cos^2\theta\right)\,.
\eea
Here $D_{h/q}(z_{h},p_{h\perp}^{2})$ and $D_{\Lambda/q}(z_{\Lambda}, p_{\Lambda\perp}^{2})$ are the unpolarized TMD FFs for $h$ and $\Lambda$, respectively. 
Meanwhile, $S(\vec{\lambda}_{\perp})$ is the soft factor, while $H(Q)$ is the hard function with the leading order expression $H^{(0)}(Q) = 1$. 
The $\vec{k}_{i \perp}$ with $i=h,\Lambda$ are the transverse momenta of the fragmenting quarks in the frame where the hadron has zero transverse momentum. 
Similarly, the $\vec{p}_{i \perp}$ are the transverse momenta of the hadrons in the frame where the fragmenting quarks have zero transverse momentum. 
These momenta are related to one another by $\vec{p}_{i \perp} = -z_{i} \vec{k}_{i\perp}$. 

It is important to realize that one could absorb part of the soft function $\sqrt{S}$ into the definition of the TMD FFs~\cite{Collins:2011zzd}. 
In this new formulation, we may rewrite the above factorization formalism in Eq.~\eqref{e.factorization.0}, so that it is of the form
\bea
\frac{d\sigma}{d\mathcal{PS} d^2 \vec{q}_\perp} =& \sigma_0 H(Q) z_\Lambda^2z_h^2 \sum_q e_q^2 
\int d^2 \vec{k}_{h \perp} d^2 \vec{k}_{\Lambda \perp} 
\nonumber \\
&\hspace{-1mm}\times\delta^{(2)}(\vec{k}_{\Lambda \perp}+\vec{k}_{h \perp} -\vec{q}_{\perp}) 
\nonumber\\
&\hspace{-1mm}\times D_{\Lambda/q}(z_\Lambda, p_{\Lambda\perp}^2; Q) D_{h/\bar{q}}(z_h, p_{h\perp}^2; Q),
\label{e.factorization}
\eea
which mimics the results of the partonic model. 
One should note that we purposely write the explicit dependence of the TMD FFs on $Q^2$, which can be derived from the usual TMD evolution formalism; see, for example,~\cite{Collins:2011zzd,Aybat:2011zv,Aybat:2011ge,Echevarria:2012pw,Echevarria:2014xaa,Kang:2015msa,Ebert:2019okf} and references therein. 
For later convenience, we define the short-hand notation
\bea
\mathcal{F}\left[D_{\Lambda/q} D_{h/\bar q}\right] &= H(Q) z_\Lambda^2 z_h^2 \sum_q e_q^2 \int d^2 \vec{k}_{h \perp} d^2 \vec{k}_{\Lambda \perp} 
\nonumber\\
& \hspace{-6mm}\times \delta^{(2)}(\vec{k}_{\Lambda \perp}+\vec{k}_{h \perp} -\vec{q}_{\perp}) 
\nonumber \\
&\hspace{-6mm} \times D_{\Lambda/q}(z_\Lambda, p_{\Lambda\perp}^2; Q) D_{h/\bar{q}}(z_h, p_{h\perp}^2; Q).
\label{e.Fcov}
\eea
With Eq.~(\ref{e.Fcov}) in hand, we can thus write the unpolarized differential cross section in the form
\bea
\frac{d\sigma}{d\mathcal{PS} d^2 \vec{q}_\perp} =& \sigma_0\, \mathcal{F} \left[D_{\Lambda/q} D_{h/\bar q}\right]\,.
\eea

When one measures the transverse polarization of the final-state $\Lambda$, one must also consider the transverse-spin dependent differential cross section, the complete expression of which was written down in~\cite{Boer:1997mf}. 
We will reproduce here the relevant terms for our analysis. 
With the short-hand notation in Eq.~\eqref{e.Fcov}, we have the expression for the transverse-spin dependent differential cross section
\bea
\frac{d\sigma(\vec{S}_\perp)}{d\mathcal{PS} d^2 \vec{q}_\perp} = & \sigma_0 \Big\{\mathcal{F} \left[D_{\Lambda/q} D_{h/\bar q}\right] + |\vec{S}_\perp| \sin(\phi_S - \phi_\Lambda)
\nonumber \\
& \times \frac{1}{z_\L M_\L} \mathcal{F}\left[\hat{\vec{P}}_{\Lambda T}\cdot \vec{p}_{\Lambda \perp} D_{1T,\Lambda/q}^{\perp}D_{h/\bar q}\right]
\nonumber\\
& + \cdots
\Big\}\,,
\label{e.spin-terms}
\eea
where $\hat{\vec{P}}_{\Lambda T} = \vec{P}_{\Lambda T}/|\vec{P}_{\Lambda T}|$ is the unit vector along the transverse momentum of the $\Lambda$ particle, as defined in Fig.~\ref{f.frame}. 
Meanwhile, $D_{1T,\Lambda/q}^{\perp}$ is the so-called \textit{polarizing fragmentation function} defined in the Trento convention~\cite{Bacchetta:2004jz} as
\bea
\hat{D}_{\L/q}& (z, \vec{p}_{\L \perp}, \vec{S}_{\perp};Q) = \frac{1}{2}\Big[ D_{\L/q}(z_{\L}, p_{\L \perp}^{2};Q) 
\nonumber \\
&+\frac{1}{ z_\L M_{\Lambda}} D_{1T, \L/q}^{\perp}(z, p_{\L \perp}^2;Q) 
\epsilon^{\rho \sigma}_{\perp} p_{\L \perp \rho} S_{\perp \sigma}\Big]\,,
\eea
where $\hat{D}_{\L/q}$ on the left-hand side can be interpreted as the number density of a polarized spin-$1/2$ hadron $\Lambda$ in an unpolarized quark, and $\vec{S}_\perp$ is its transverse polarization vector. 

In trying to connect the theoretical formalism above with the BELLE collaboration's experimental measurement of $\Lambda$ polarization, one encounters several subtleties.

First is the direction with respect to which BELLE measures $\Lambda$ polarization. 
Defining $\vec{m} = -\hat{\vec{P}}_h$, with $\hat{\vec{P}}_h$ ($\hat{\vec{P}}_{\L}$) the unit vector along the momentum of the hadron $h$ (the $\Lambda$), we see that BELLE measures $\Lambda$ polarization along the direction $\hat{\vec{n}} \propto \vec{m}\times \hat{\vec{P}}_{\L}$, perpendicular to the hadronic plane in Fig.~\ref{f.frame}.
On the other hand, the polarization vector $\vec{S}_\perp$ in the above formalism is transverse with respect to the leptonic plane in Fig.~\ref{f.frame}. 
Because of this, we need to perform an additional projection onto the $\hat{\vec{n}}$-direction. 

Second of all, there are additional terms as denoted by ``$\cdots$'' in Eq.~\eqref{e.spin-terms}~\cite{Boer:1997mf}. 
One such term involves a convolution of transversity FFs $H_{1, \Lambda/q}(z_\Lambda, p_{\L \perp}^2;Q)$ for the $\Lambda$ hadron with the Collins FFs $H_{1, h/\bar q}^{\perp}(z, p_{h\perp}^2;Q)$ for the light hadron $h$. 
Such a term has an azimuthal dependence of $\sin(\phi_S + \phi_\Lambda)$. 
In principle, the optimal strategy to isolate, and thus extract unambiguously, the PFFs $D_{1T, \L/q}^{\perp}$ would be to measure and disentangle all of these different azimuthal dependencies, just like in the usual SIDIS spin measurements~\cite{Bacchetta:2006tn}. 
This has not yet been done by the BELLE collaboration. 
Surprisingly, though, if one integrates over $\vec{q}_\perp$ in the formalism, all the other terms vanish and we are left with only the term involving the PFF $D_{1T, \L/q}^{\perp}$ for the spin-dependent cross section~\footnote{We thank D.~Boer and H. Matevosyan for very insightful communication concerning this point.}. 

Since the experimental data are expressed only as a function of $z_\Lambda$ and $z_h$, and are inclusive over $\vec{q}_\perp$, our analysis of the experimental data to extract the PFFs is thus justified. 
Eventually with the transverse momentum integrated, the measured $\Lambda$ polarization denoted as $P_{\perp}^\Lambda$ will be given by
\bea
P_{\perp}^\Lambda(z_\L, z_h) =  \left.\frac{d\Delta\sigma(\vec{S}_\perp)}{d\mathcal{PS}} \right/ \frac{d\sigma}{d\mathcal{PS}}\,,
\label{e.pol}
\eea
where $\Delta \sigma(\vec{S}_\perp) = \left[\sigma(\vec{S}_\perp) - \sigma(-\vec{S}_\perp)\right]/2$, and the denominator is the unpolarized cross section. 
%

\section{Fit Results and Predictions}
\label{s.Fit Results}
In this section, we first provide the parametrization used for the extraction of polarizing fragmentation functions, and give an expression for the asymmetry $P_{\perp}^\Lambda(z_\L, z_h)$ within our model. We then describe our fitting procedure and the fitted results. Finally, we make a prediction for the $\Lambda$ polarization in semi-inclusive deep inelastic scattering. 

\subsection{Fitting scheme}
\label{ss.fitscheme}
All available data are measured at the same hard scale $Q=10.58$~GeV at the BELLE experiment; thus, TMD evolution for the relevant TMD FFs is not needed. 
Because of this, we can model these TMD FFs using simple Gaussians and extract them at this particular scale $Q$. 
We model the unpolarized TMD FFs as Gaussians
\bea
D_{h/q}(z_{h},p_{h\perp}^{2};Q) &= D_{h/q}(z_{h},Q) \; \frac{e^{-p_{h \perp}^{2}/\langle p_{h \perp}^{2}\rangle}}{\pi \langle p_{h \perp}^{2} \rangle},
\\
D_{\Lambda/q}(z_{\Lambda},p_{\Lambda\perp}^{2};Q) &=  D_{\Lambda/q}(z_{\Lambda},Q)\; \frac{e^{-p_{\Lambda \perp}^{2}/\langle p_{\Lambda \perp}^{2}\rangle}}{\pi \langle p_{\Lambda \perp}^{2} \rangle},
\label{eq.ff_upol}
\eea
where we take $\langle p_{h\perp}^{2}\rangle=0.19$ GeV$^{2}$ from~\cite{Anselmino:2014pea} for the light hadrons $h$. For $\L$, we assume $\langle p_{\Lambda \perp}^{2} \rangle = \langle p_{h \perp}^{2} \rangle$ in this paper. 
We model the polarizing fragmentation functions $D_{1T,\Lambda/q}^{\perp}$ according to the equation
\bea
\hspace{-2mm}D_{1T,\Lambda/q}^{\perp}(z_{\Lambda},p_{\Lambda \perp}^2;Q) = D_{1T,\Lambda/q}^{\perp}(z_{\Lambda},Q)
\frac{e^{-p_{\Lambda \perp}^{2}/\langle M_{D}^{2}\rangle}}{\pi \langle M_{D}^{2} \rangle}.
 \label{eq.ff_pol}
\eea
Here we write the polarized collinear function $D_{1T,\Lambda/q}^{\perp}(z_{\Lambda}, Q)$ simply as a modulation of the unpolarized collinear function $D_{\Lambda/q}(z_{\Lambda}, Q)$ by an additional collinear function $\mathcal{N}_{q}(z_{\L})$
\bea
D_{1T,\Lambda/q}^{\perp}(z_{\Lambda},Q) = \mathcal{N}_{q}(z_{\Lambda}) D_{\Lambda/q}(z_{\Lambda},Q)\,,
\eea
and we parametrize $\mathcal{N}_{q}(z_{\Lambda})$ by the formula
\bea
\mathcal{N}_{q}(z_{\Lambda})= N_{q}z_{\Lambda}^{\alpha_{q}}(1-z_{\Lambda})^{\beta_{q}} \frac{(\alpha_q+\beta_q-1)^{\alpha_q+\beta_q-1}}{(\alpha_q-1)^{\alpha_q-1}{\beta_q}^{\beta_q}}.
\eea
The Gaussian width $\langle M_{D}^{2} \rangle$ differs from the unpolarized width $\langle p_{\Lambda \perp}^{2} \rangle$ by an auxiliary width $M_1$ obeying the equality~\cite{Anselmino:2013vqa,Gamberg:2013kla}
\bea
    \langle M_{D}^{2}\rangle  \equiv\bigg(\frac{1}{\langle  p_{\L \perp}^{2} \rangle}+\frac{1}{M_{1}^{2}}\bigg)^{-1}=\frac{ M_{1}^{2}\langle p_{\L\perp}^{2}\rangle}{M_{1}^{2}+\langle p_{\L\perp}^{2}\rangle},
\eea
from which it is clear that $M_{1}$ characterizes the scale of spin corrections to $\langle p_{\Lambda\perp}^{2}\rangle$.
We choose to fit $\langle M_{D}^{2} \rangle$ -- of course, $M_1$ can be easily determined once $\langle M_{D}^{2} \rangle$ is known. 

In order to maintain the interpretation of the spin-dependent fragmentation functions $\hat{D}_{\L/q}(z, \vec{p}_{\L \perp}, \vec{S}_{\perp};Q)$ as probability densities, the positivity bound
\begin{align}
\frac{p_{\Lambda\perp}}{z_{\Lambda} M_\Lambda}\left|D_{1T, \Lambda/q}^{\perp}(z_{\Lambda},p_{\Lambda\perp}^{2};Q)\right|\leq D_{\Lambda/q}(z_{\Lambda},p_{\Lambda\perp}^{2};Q)
\end{align}
given in~\cite{Bacchetta:1999kz,Metz:2016swz}, must be satisfied. 
We thus implement the fit constraints
\bea
   & \alpha_q > 1\,,~ \beta_q > 0\,,~
   \langle M_D^2\rangle < \langle p_{\L\perp}^{2}\rangle\,,
   \\
      & |N_{q}|\leq \sqrt{2e} \frac{\langle M_{D}^{2}\rangle}{\langle p_{\L\perp}^{2}\rangle}\frac{M_{\L}}{M_{1}}\,,
\eea
which are sufficient conditions for the enforcement of the positivity bound. 
Moreover, it is useful to define the $p_{\L \perp}^2$-moment of the TMD PFFs
\bea
D_{1T, \L/q}^{\perp (1)}(z_\L, Q) &\equiv \int d^2\vec{p}_{\L \perp}
\frac{p_{\L \perp}^2}{2z_\L^2 M_\L^2} D_{1T, \L/q}^{\perp}(z_\L, p_{\L \perp}^2; Q)
\nonumber\\
&=\frac{\langle M_D^2\rangle}{2z_\L^2 M_\L^2} D_{1T,\Lambda/q}^{\perp}(z_{\Lambda},Q).
\label{e.moment}
\eea

Using our parametrization, all momenta can be integrated out analytically, so that the cross sections take on the forms
\begin{align}
\hspace{-4mm}\frac{d\sigma}{d\mathcal{PS}} 
    =&\sigma_{0}H(Q)\sum_{q}e_{q}^{2} D_{\Lambda/q}(z_{\Lambda}, Q) D_{h/\bar{q}}(z_h, Q),
\\
\hspace{-4mm} \frac{d\Delta \sigma(\vec{S_\perp})}{d \mathcal{PS}} 
    =&\sigma_{0} H(Q) \frac{z_h\sqrt{\pi}}{2z_{\L}}\frac{\langle M_{D}^{2}\rangle}{M_{\Lambda}\sqrt{z_{h}^{2}\langle M_{D}^{2}\rangle+z_{\Lambda}^{2}\langle p_{h\perp}^{2}\rangle}}
    \nonumber\\
&  \times\sum_{q}e_{q}^{2}D_{1T,\Lambda/q}^{\perp}(z_{\Lambda},Q)D_{h/\bar{q}}(z_{h},Q).
\end{align}
As such, we finally obtain the following expression for the $\Lambda$ polarization $P_{\perp}^{\Lambda}(z_\L, z_h)$ from Eq.~\eqref{e.pol},
\bea
P_{\perp}^{\Lambda}(z_\L, z_h) =&\frac{z_h\sqrt{\pi}}{2z_{\L}}\frac{\langle M_{D}^{2}\rangle}{M_{\Lambda}\sqrt{z_{h}^{2}\langle M_{D}^{2}\rangle+z_{\Lambda}^{2}\langle p_{h\perp}^{2}\rangle}}
\nonumber\\ 
&\times \frac{\sum_{q}e_{q}^{2}D_{1T,\Lambda/q}^{\perp}(z_{\Lambda},Q)D_{h/\bar{q}}(z_{h},Q)}{\sum_{q}e_{q}^{2}D_{\Lambda/q}(z_{\L},Q)D_{h/\bar{q}}(z_{h},Q)}.
\label{e.polee}
\eea

To compute $P_{\perp}^{\Lambda}(z_\L, z_h)$, we use the AKK08 parametrization \cite{Albino:2008fy} of the collinear $\Lambda$ fragmentation functions. 
Currently, there are no available collinear fragmentation functions which separate the $\L$ and $\bar{\L}$ contributions. 
While the work in \cite{Anselmino:2019acqd} took $D_{\bar{\L}/q} = D_{\L/\bar{q}} = 0$ with $q=u,d,s$, this scheme does not adequately describe $\L+h$ production. 
For example, in the
$e^- + e^+\to \Lambda+\pi^++X$ process, one of the dominant contributions to the cross-section is given by the $D_{\L/\bar{u}}(z_\L, Q)$. 
Since the work in \cite{Anselmino:2019cqd} neglected all sea quark contributions, this would lead to a very small asymmetry, which conflicts with the BELLE data. %
For this paper, we assume $D_{\L/q}=D_{\bar\L/q}=\frac{1}{2}D_{\L/\bar{\L}\leftarrow q}$ for all quark flavors. 

For the fragmentation functions of pions, we choose the DSS14 parametrization given in~\cite{deFlorian:2014xna}, which is an update of the previous DSS07 fragmentation functions~\cite{deFlorian:2007aj}. %
As such an update is not available for kaons, we choose the DSS07 parametrizations for the fragmentation functions of kaons. 

In order to fit the non-perturbative TMD PFFs $D_{1T,\Lambda/q}^{\perp}(z_{\Lambda},p_{\L\perp}^2; Q)$, we use the typical flavor-dependent parameters $N_{q},~\alpha_{q}$, and $\beta_{q}$, similar to the parametrization used in~\cite{Echevarria:2014xaa} for the Sivers functions. %
In this paper, for the polarization of the $\L$, we fit the 11 parameters $N_u$, $N_d$, $N_s$, $N_{\mathrm{sea}}$, $\alpha_{u}$, $\alpha_{d}$, $\alpha_{s}$, $\alpha_{\mathrm{sea}}$, $\beta_{\mathrm{val}}$, $\beta_{\mathrm{sea}}$, and $\langle M_{D}^{2}\rangle$. 
The parameters labeled sea apply to the remaining considered flavors, namely $\bar{u}$, $\bar{d}$, and $\bar{s}$. 
Furthermore, in order to fit the $\bar{\L}$ polarization, we take $D^{\perp}_{1T,\bar{\L}/\bar{q}}(z_{\Lambda},p_{\L\perp}^2;Q) = D^{\perp}_{1T,\L/q}(z_{\Lambda},p_{\L\perp}^2;Q)$, by invariance under charge conjugation.

Following Ref.~\cite{Pisano:2018skt}, we use a bootstrap method to generate the uncertainty band for the PFF and polarization alike. 
For this purpose, we generate 200 replicas; to generate one replica, we shift the reported polarization associated with each data point by Gaussian noise with standard deviation equal to the experimental error. 
The fit is performed on the noisy data, resulting in a set of parameters. 
We perform this 200 times to obtain 200 fits, from each of which we calculate $P_{\perp}^{\L}$ (or the PFF). 
The middle 68$\%$ of these values are selected point-by-point. 
At each point, the minimum and maximum of this middle 68$\%$ are considered to be the upper and lower errors.
%

\subsection{Fit Results}
We use the \texttt{MINUIT} package~\cite{James:1975dr} from \texttt{CERNLIB} to perform the fit. 
The parameters as well as the $\chi^2/d.o.f$ of the fit are presented in \tref{params}. 
The $\chi^2/d.o.f$ of 1.694 suggests that the fit is of reasonably good quality. 
One must note that we have restricted ourselves to fit the experimental data with $z_{h}<0.5$, for a total of 96 data points. 
It is also important to note that when these parameters are used to describe the data globally, without removing the $z_{h}>0.5$ data, we have $\chi^2/d.o.f = 2.421$. 
This could indicate a sizable contribution of threshold logarithms~\cite{Anderle:2012rq} and target mass corrections~\cite{DeRujula:1976baf,Accardi:2008pc,Guerrero:2015wha} in this region. 

While the advertised $\chi^2/d.o.f$ is $1.694$, a large contribution of the $\chi^2$ comes from two ``problematic'' points, the point at $z_h = 0.243$, $z_\L = 0.35$ for the $\L+K^+$ process and the point at $z_h = 0.245$, $z_\L = 0.35$ for the $\bar{\L}+K^-$ process. 
If the $\chi^2$ contributions from these points are removed, the $\chi^2/d.o.f$ becomes $1.499$. In fact removing these points from the fitting procedure altogether leads to a $\chi^2/d.o.f$ of $1.180$. 
In the future, it would be interesting to investigate these two points in  more detail.

\begin{figure*}[htb!]
    \centering
    \includegraphics[width=0.9\textwidth,trim={10mm 0mm 20mm 15mm},clip]{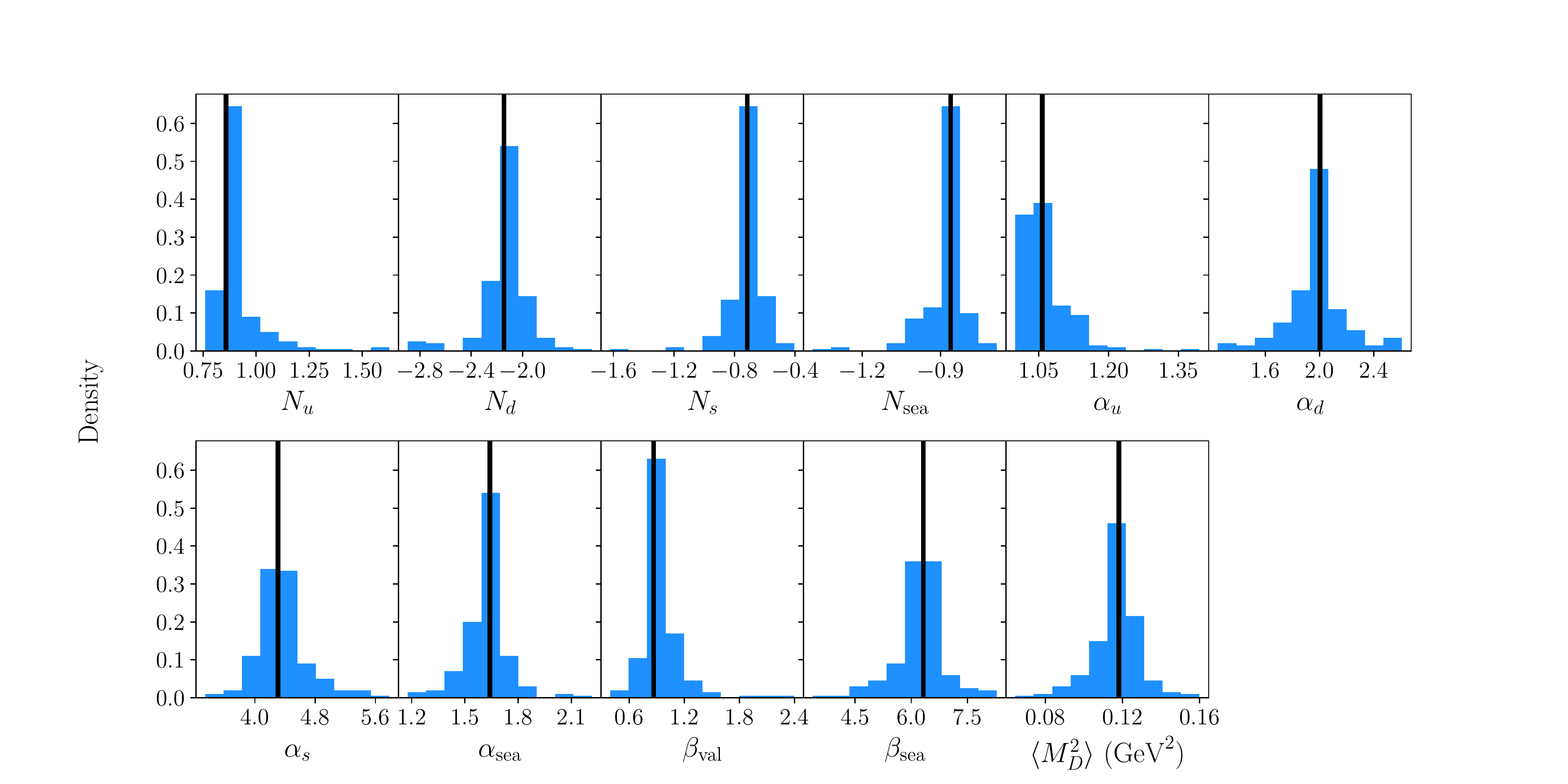}
    \caption{Distributions of \texttt{MINUIT} parameters from 200 replicas. The black lines represent the parameter values which are determined from the best fit of the actual experimental data. Each histogram is normalized such that the heights of its bars sum to unity. }
\label{f.hists}
\end{figure*}

\fref{hists} contains histograms of the distributions of fit parameters, which are determined by the fits to the replicated data sets. 
We find that the modes of the histograms agree well with the determined values of the central fit. 
This agreement indicates that the values of the parameters are well-constrained, and not appreciably sensitive to variations of the central point within the experimental uncertainties.

\begin{figure*}[htb!]
     \centering
     \includegraphics[width=0.9\textwidth,trim={15mm 0mm 30mm 11mm},clip]{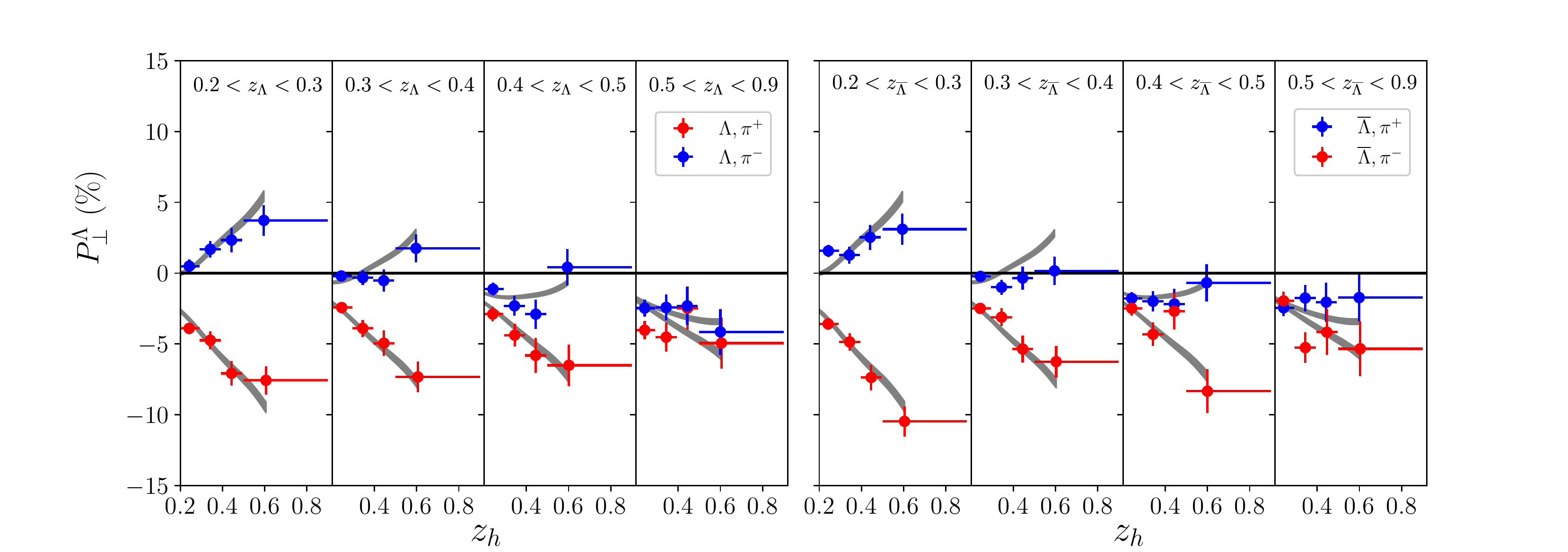}
     \caption{The fit to the experimental data for $\pi$ mesons is shown, with the gray uncertainty band displayed is generated by the replicas at 68$\%$ confidence. The left plots are for the production of $\L+\pi^\pm$, while the right plots are for the production of $\bar{\L}+\pi^\pm$.}
   \label{f.fitplusdata_pi}
\end{figure*}
\begin{figure*}[htb!]
     \centering
     \includegraphics[width=0.9\textwidth,trim={15mm 0mm 30mm 11mm},clip]{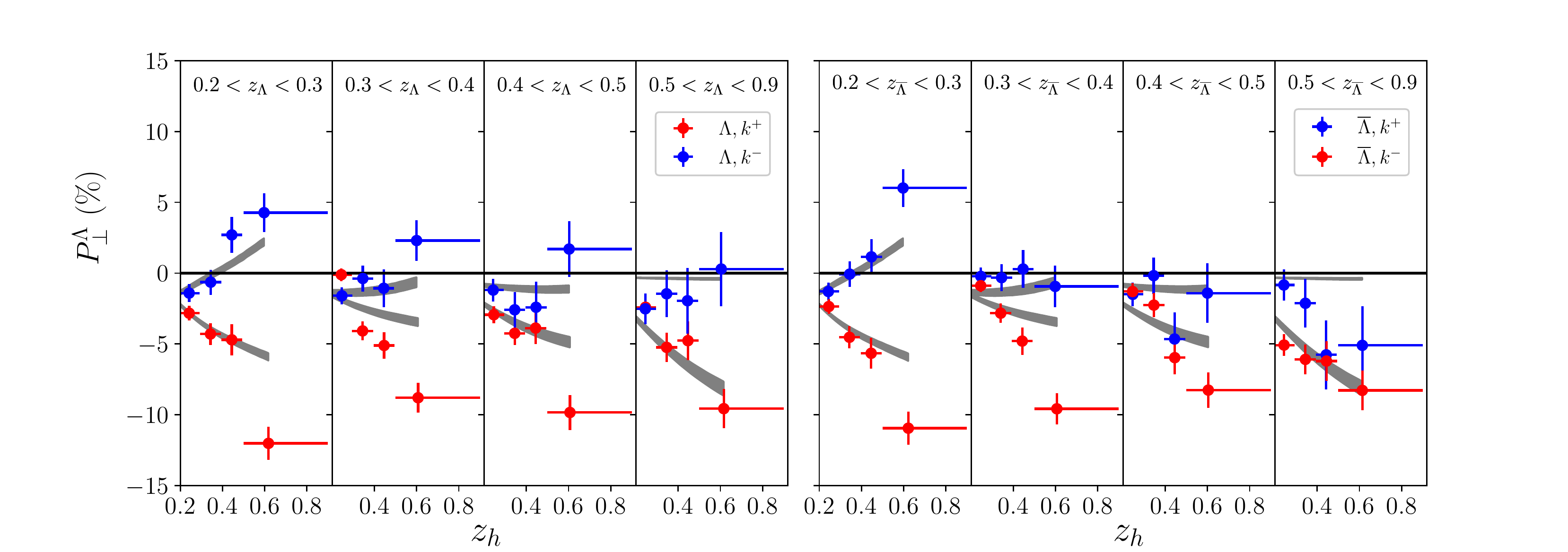}
     \caption{Same as \fref{fitplusdata_pi} but for the production of $\L + K^\pm$ (left) and $\bar\L + K^\pm$ (right).}
   \label{f.fitplusdata_k}
\end{figure*}
%

In Figs.~\ref{f.fitplusdata_pi} and \ref{f.fitplusdata_k} we plot the experimental data, as well as the result of our fit for the $\L$ polarization $P_\perp^\L$ in the back-to-back production of $\Lambda (\bar{\L})+\pi^\pm$ and $\Lambda (\bar{\L})+K^\pm$, respectively. 
The gray uncertainty bands displayed are generated by the replicas at 68$\%$ confidence. 
For \fref{fitplusdata_pi}, the left plots correspond to $\L+\pi^\pm$ production, while the right plots correspond to $\bar{\L}+\pi^\pm$ production. 
Likewise, the left (right) plots are for the $\L$ ($\bar{\L}$) production associated with $K^\pm$. 
One should note that the data points with $z_{h}>0.5$ are not included in our fit, and thus we see that the global comparison with our theoretical results is of slightly lower quality. 
We further observe that our model seems to describe the $\L (\bar{\L})+\pi^\pm$ data better than the $\L (\bar{\L})+K^\pm$ data; indeed, we find $\chi^{2}/ndata=1.223$ for pions, and $1.802$ for kaons.

\bef    
\includegraphics[width= 0.4\paperwidth, trim ={0mm 2mm 0mm 3mm},clip]{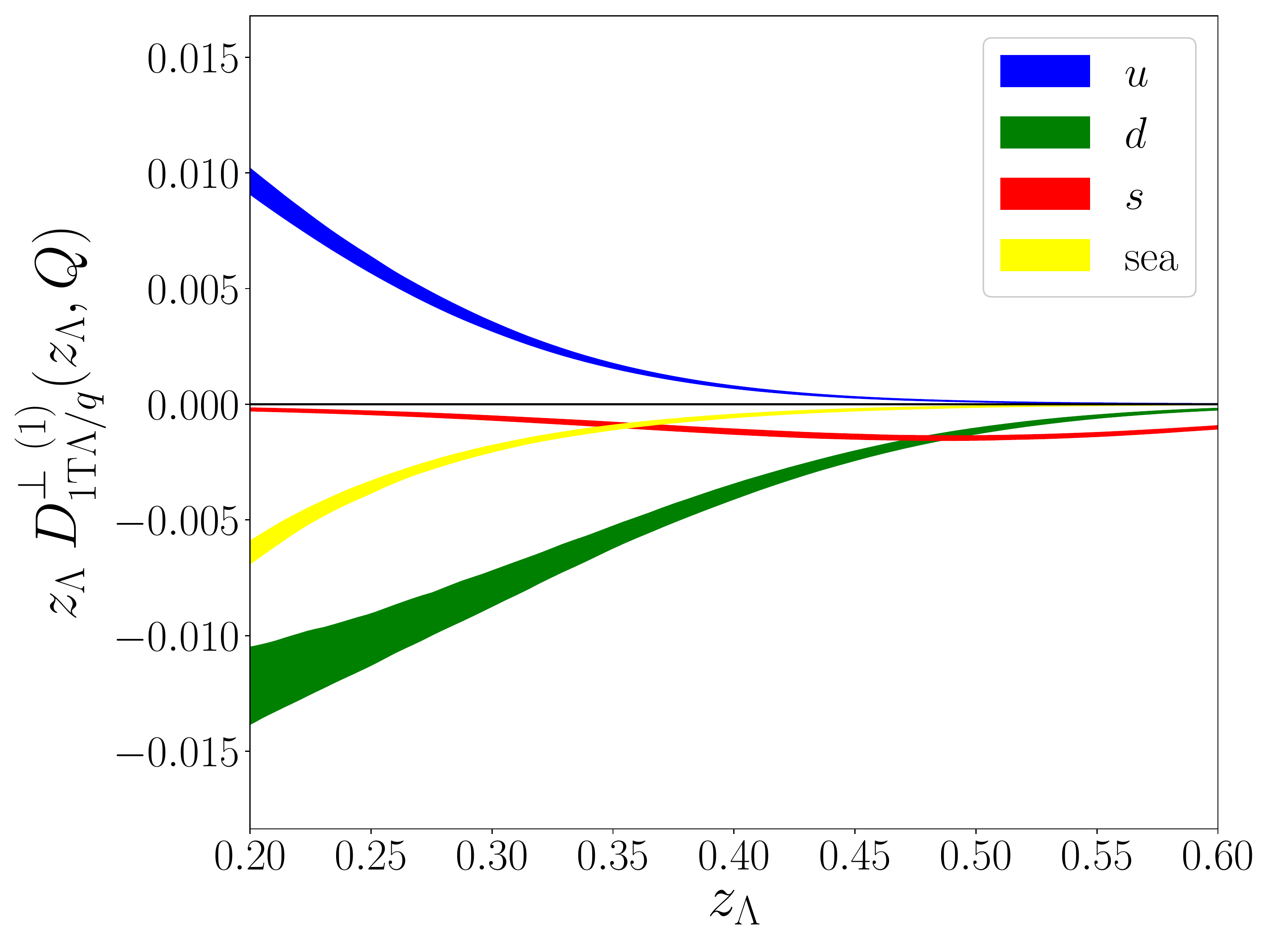}
    \caption{The polarizing fragmentation functions $z_\L D_{1T, \L/q}^{\perp(1)}(z_\L, Q)$, defined in Eq.~\eqref{e.moment}, are plotted as functions of $z_{\Lambda}$ for different quark flavors, at 68$\%$ confidence.}
\label{f.PFF}
\end{figure}

\begin{table}[ht!] 
\begin{ruledtabular}
\begin{tabular}{ll}
\multicolumn{2}{c}{
$\chi^{2}/d.o.f =1.694$} \\[5pt]
\hline
$N_{u}=0.858_{-0.011}^{+0.108}$ & $N_{d}=-2.144_{-0.088}^{+0.156}$ \\[5pt] 
$N_{s}=-0.716_{-0.068}^{+0.070}$& $N_{\mathrm{sea}} = -0.861_{-0.086}^{+0.026}$\\[5pt]
$\alpha_{u} = 1.058_{-0.044}^{+0.050}$ & $\alpha_{d}=2.004_{-0.196}^{+0.123}$\\[5pt]
$\alpha_{s} = 4.306_{-0.185}^{+0.326}$&
$\alpha_{\mathrm{sea}} = 1.641_{-0.102}^{+0.053}$\\[5pt]
$\beta_{\mathrm{val}}=0.866_{-0.046}^{+0.218}$ & $\beta_{\mathrm{sea}} = 6.325_{-0.522}^{+0.240}$\\[5pt]
$ \langle M_{D}^{2} \rangle = 0.118_{-0.012}^{+0.007}\mathrm{~GeV}^{2}$
\end{tabular}
\end{ruledtabular}
\caption{\label{t.params} Listed are the parameter values with uncertainties. The central values are taken from the fit with the actual BELLE data~\cite{Guan:2018ckx} (no Gaussian noise), while the uncertainties are calculated from the middle 68$\%$ of parameter values generated from 200 replicas (see the discussion in Sec. \ref{ss.fitscheme}).}
\end{table}

In \fref{PFF}, we plot $z_{\L}D_{1T,\Lambda/q}^{\perp(1)}(z_{\Lambda}, Q)$, defined in Eq.~\eqref{e.moment}, as a function of $z_{\L}$ for $u$, $d$, $s$ and sea quarks, at 68$\%$ confidence. 
We find that the PFF for the $u$ quark is positive, while those of the $d$ and $s$ quarks are negative. 
We also find a sizable negative sea quark contribution. 
These signs are consistent with the qualitative analysis in the BELLE experimental paper~\cite{Guan:2018ckx}. 
In terms of the magnitude of the PFFs, we find that the $u$ and $d$ quarks are comparable, while the PFF for the $s$ quark is smaller by almost an order of magnitude, and it plays a more important role in the relatively large $z_\L\gtrsim 0.4$. 
The PFFs for sea quarks are sizable mostly in the relatively small $z_\L \lesssim 0.3$ region.

One can understand these findings qualitatively. 
For example, the $\L+\pi^-$ processes are dominated by the contribution of $D_{1T,\L/u}^{\perp} D_{\pi^-/\bar u}$ in Eq.~\eqref{e.polee}. 
As this subset of BELLE data has large positive $\Lambda$ polarization ($z_\L \lesssim 0.4$), we find that the sign of the $u$-quark PFF is positive. 
Likewise, the $\L+\pi^+$ processes are dominated by the contribution of $D_{1T,\L/d}^{\perp} D_{\pi^+/\bar d}$. 
Due to the large negative polarization, we find that the sign of the $d$-quark PFF is negative. 
Finally the $\L+K^+$ process is dominated by the contribution of $D_{1T,\L/s}^{\perp} D_{K^+/\bar s}$. 
We then determine the sign of the $s$-quark PFF to be negative, although our best fit gives a very small PFF for the $s$-quark. 
Finally the sea quarks usually play more important roles in the relatively small $z_\Lambda$ region. 
In this set of BELLE data, it starts to become more important for $z_\Lambda \lesssim 0.3$. 
We find negative PFFs for sea quarks, which are smaller in size compared with those for $u$ and $d$ quarks.

\subsection{Predictions for the SIDIS process}
We now present a phenomenological prediction for the polarization of the $\Lambda$ particle, produced in the SIDIS process, $e(\ell)+p(P)\rightarrow e(\ell^\prime)+\L(P_{\L},\vec{S}_{\perp})+X$. 
As emphasized in~\cite{Boer:2010ya}, the measurement of $\Lambda$ polarization in SIDIS furnishes an experimental verification of the universality of the TMD PFF $D_{1T}^{\perp}(z_\Lambda, p_{\L \perp}^2; Q)$, which has been predicted to be the same as those measured in $e^+e^-$ annihilation~\cite{Metz:2002iz,Collins:2004nx,Meissner:2008yf,Boer:2010ya}. 
We define the standard SIDIS variables
\bea
x_B = \frac{Q^2}{2P\cdot q},\qquad
y = \frac{P\cdot q}{P\cdot \ell},\qquad
z_\Lambda = \frac{P\cdot P_\Lambda}{P\cdot q},
\eea
where $Q^2 = -q^2 = - (\ell' - \ell)^2$. 
The differential cross section is given by
\bea
\frac{d\sigma(\vec{S}_\perp)}{d\mathcal{PS} d^2 \vec{P}_{\Lambda T}} = & \sigma_0^{\rm DIS} \Big\{\mathcal{F} \left[f_{q/p}D_{\Lambda/q} \right] + |\vec{S}_\perp| \sin(\phi_S - \phi_\Lambda)
\nonumber \\
&\times \frac{1}{z_\L M_\L}\mathcal{F}\left[\hat{\vec{P}}_{\Lambda T}\cdot \vec{p}_{\Lambda \perp} f_{q/p}D_{1T,\Lambda/q}^{\perp}\right]\Big\}\,,
\label{e.sidis}
\eea
where for SIDIS we have the phase-space element $d\mathcal{PS} = dx_B\, dy\, dz_\Lambda$, the usual unpolarized TMD PDFs $f_{q/p}(x_B, k_\perp^2; Q)$, and the leading-order scattering cross section
\bea
\sigma_0^{\mathrm{DIS}} = \frac{2\pi\alpha_{\mathrm{em}}^2}{Q^2}\frac{1+(1-y)^2}{y}.
\eea
Collecting the results above, we find that the convolution for SIDIS is
\bea
\mathcal{F}\left[f_{q/p}D_{\Lambda/q} \right] &= H^{\rm DIS}(Q) \sum_q e_q^2 \int d^2 \vec{k}_{\perp} d^2 \vec{p}_{\Lambda \perp} 
\nonumber\\
& \hspace{-6mm}\times \delta^{(2)}(z_\Lambda \vec{k}_{\perp}+\vec{p}_{\Lambda \perp} -\vec{P}_{\Lambda T})
\nonumber \\
&\hspace{-6mm} \times f_{q/p}(x_B, k_{\perp}^2; Q) D_{\Lambda/q}(z_\Lambda, p_{\Lambda\perp}^2; Q),
\label{e.Fcov-dis}
\eea
\bef
    \includegraphics[width=0.4\paperwidth]{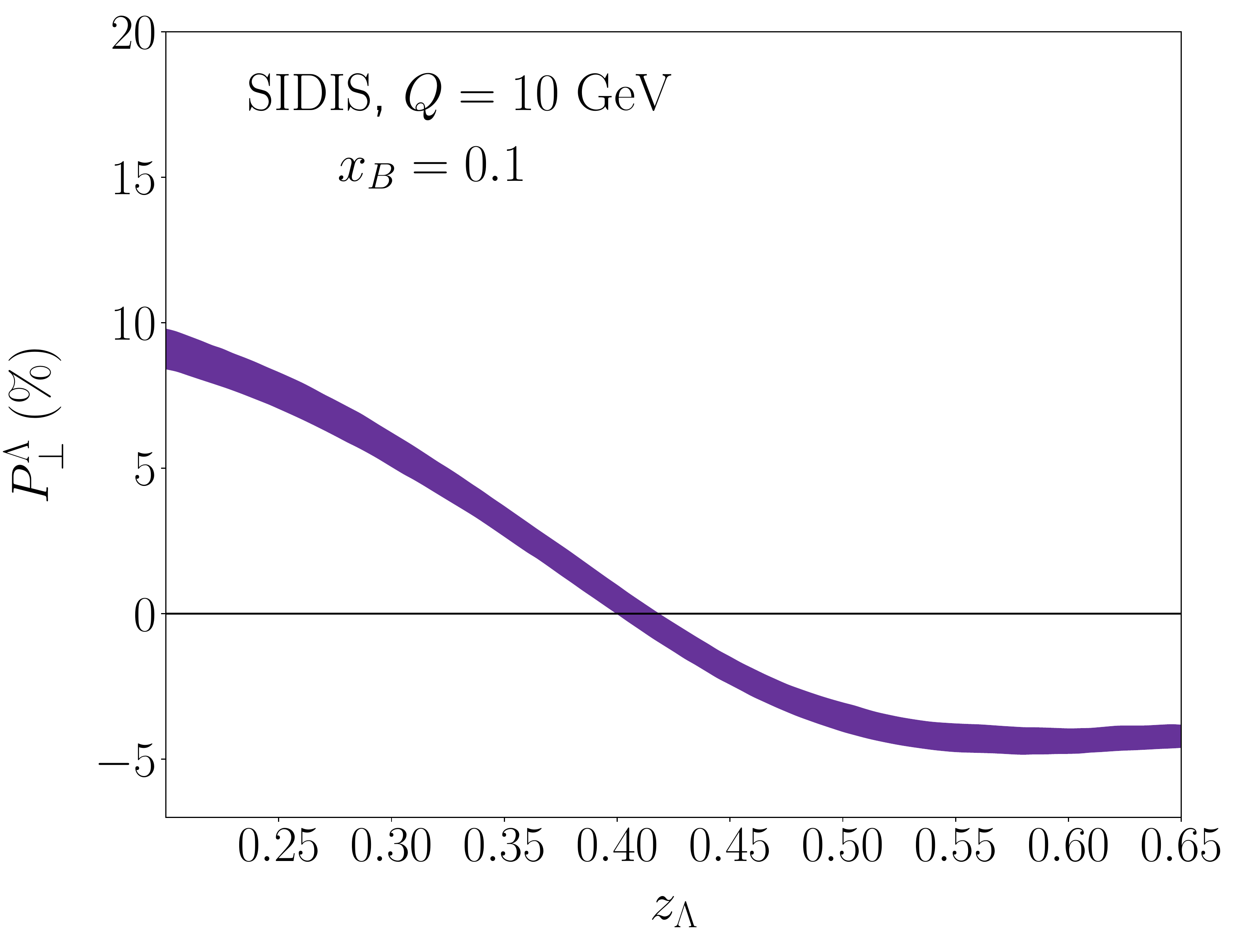}
    \caption{Our prediction of the transverse $\Lambda$ polarization $P_{\perp}^\Lambda(x_B, z_\Lambda)$ in SIDIS is plotted as a function of $z_\L$ for typical values of the kinematic variables $Q = 10$ GeV and $x_{B} = 0.1$ at the EIC. The uncertainty band is generated at 68$\%$ confidence.}
\label{f.plot4}
\end{figure}
where $H^{\rm DIS}(Q)$ is the hard function for SIDIS, with $H^{\rm DIS(0)}(Q) = 1$ at leading order. 
Meanwhile, $\vec{P}_{\L T}$ is the transverse momentum of the final-state $\L$, $\vec{k}_{\perp}$ is the transverse momentum of the quark relative to the initial-state parent proton and $\vec{p}_{\Lambda \perp}$ is the transverse momentum of the final-state $\Lambda$ with respect to the fragmenting quark.

\bef
    \includegraphics[width=0.4\paperwidth]{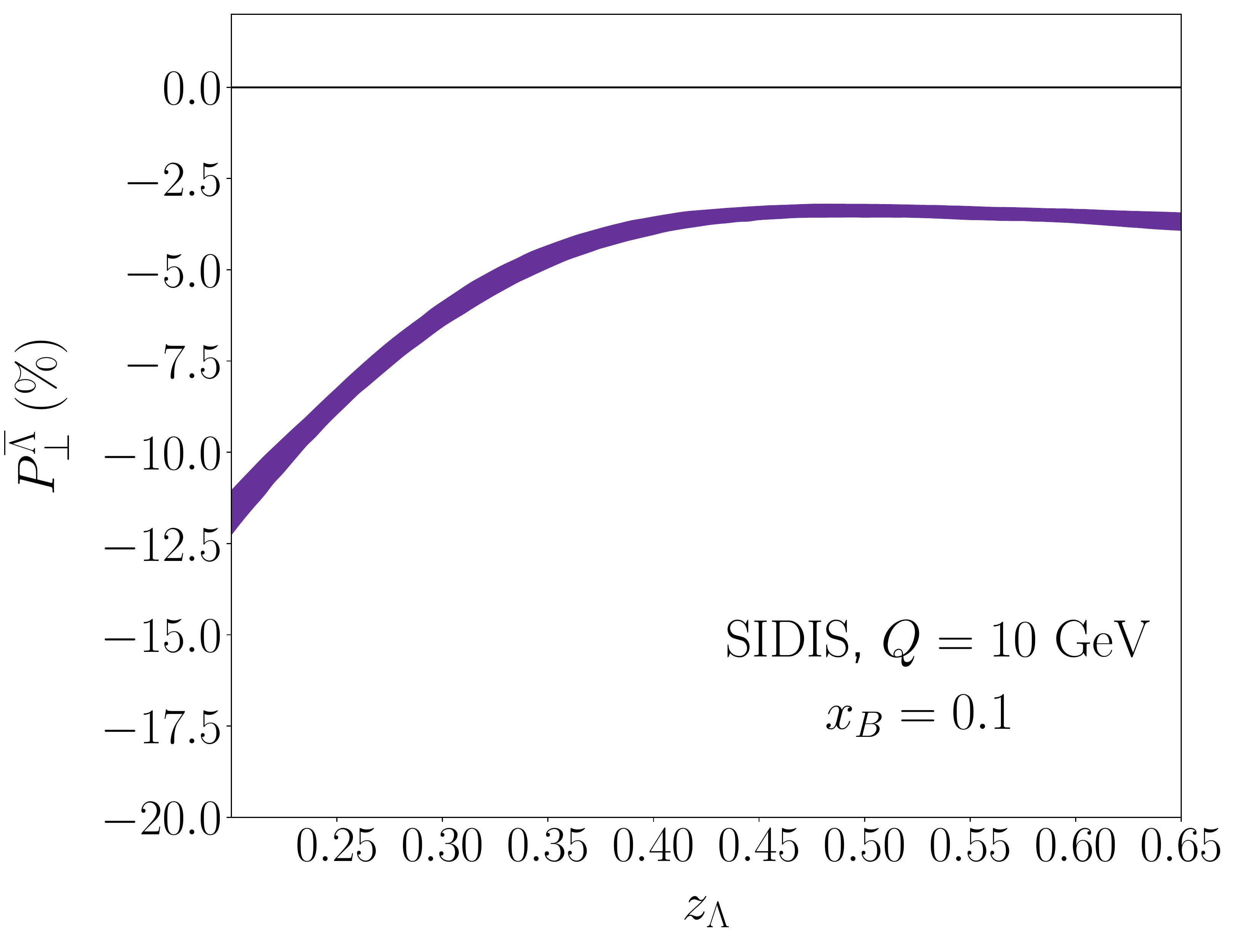}
    \caption{Same as \fref{plot4} but for $\bar{\L}$ production.}
\label{f.plot5}
\end{figure}

For our calculations, we use the parametrizations of $\Lambda$ fragmentation functions in  Eqs.~(\ref{eq.ff_upol}) and (\ref{eq.ff_pol}) from the previous section. 
For the unpolarized TMD PDFs, we use the parametrization
\begin{align}
f_{q/p}(x_{B},k_{\perp}^2; Q)=f_{q/p}(x_{B},Q)\frac{e^{-k_{\perp}^{2}/\langle k_{\perp}^{2} \rangle}}{\pi\langle k_{\perp}^{2}\rangle},
\label{eq.pdfpar}
\end{align}
with $\langle k_{\perp}^{2}\rangle=0.61$ GeV$^{2}$, as extracted in \cite{Anselmino:2014pea}. 
Using Eq.~(\ref{eq.pdfpar}) and integrating over $\vec{P}_{\Lambda T}$, we find that the $\Lambda$ polarization has the analytic form 
\begin{align}
P_{\perp}^{\L}(x_B, z_\L) =&
\frac{\sqrt{\pi}}{2z_{\L}}
\frac{ \langle M_D^2 \rangle}{ M_{\L} \sqrt{\langle M_D^2 \rangle+z_{\L}^2\langle k_{\perp}^2 \rangle}}
\nonumber\\
& \times \frac{\sum\limits_{q}e_{q}^{2}f_{q/p}(x_{B},Q)D_{1T,\Lambda/q}^{\perp}(z_{\Lambda},Q)}{\sum\limits_{q}e_{q}^{2}f_{q/p}(x_{B},Q)D_{\L/q}(z_{\L},Q)}.
\end{align}

In \fref{plot4}, we plot the transverse polarization as a function of $z_{\L}$ at $Q=10$ GeV and $x_B = 0.1$, which are consistent with the typical kinematics at the future Electron Ion Collider (EIC)~\cite{Boer:2011fh,Accardi:2012qut,Aschenauer:2017jsk,Aidala:2020mzt}.
We have used the \texttt{CT14lo} collinear PDFs given in~\cite{Dulat:2015mca}. 
To generate the uncertainty band, we use the 200 sets of fitted parameters and plot the band generated from the middle 68$\%$. We predict an asymmetry of roughly 10$\%$ for $\Lambda$ production in this kinematic region. 
As the size of the asymmetry is on par with that of other single-spin asymmetries, this measurement should be feasible at the EIC. 
We note that the sign of the polarization is due to the interplay between the contributions of $u$ and $d$ quarks. 
At small $z_{\L}\lesssim 0.4$ the magnitudes of the $u$-quark and $d$-quark PFFs are similar. 
However the contribution from the $u$ quark is weighted by the much larger fractional electric charge $e_u^2=4/9$ vs $e_d^2=1/9$, resulting in a positive asymmetry. 
At large $z_{\L}\gtrsim 0.4$ the magnitude of the $d$-quark PFF is much larger and wins over the enhancement from the electric charges, leading to a negative asymmetry.
In \fref{plot5}, we plot our prediction for $\bar{\L}$ production at $x_B = 0.1$. 
The magnitude and sign of the polarization can be interpreted by noting that at $x_B = 0.1$ the $u$ and $d$-quark PDFs are the dominant contributions. 
However, since $u$ and $d$-quarks are sea quarks of $\bar{\L}$ and the sea quark PFFs are negative, we thus have a negative polarization. 
The magnitude of the asymmetry is roughly $-10\%$ at small $z_{\L}$ and gradually decreases in size to be around $-5\%$ as $z_{\L}$ increases. 
This is consistent with the behavior of sea quark PFFs, which decreases in size as $z_\L$ increases. 
Our analysis indicates that the polarizations of $\Lambda$ and $\bar \Lambda$ in SIDIS could serve as good observables for the extraction of valence and sea quark PFFs. 

\section{Conclusions}
\label{s.Conclusions} 
In this paper, we have demonstrated that transverse momentum dependent polarizing fragmentation functions (TMD PFFs) can be extracted from the polarization measurement for both $\Lambda$ and $\bar{\Lambda}$ at BELLE in~\cite{Guan:2018ckx}. 
In the measurement, $\Lambda~(\bar{\Lambda})$ and a light hadron (pion or kaon) are produced in the back-to-back configuration, and a TMD factorization formalism can thus be applied to analyze the experimental data. 
As all of the experimental data from this single measurement were collected at the same scale, no TMD evolution is needed. We thus perform an extraction of the TMD PFFs using a simple Gaussian model. The resulting PFFs are constrained by the BELLE data to be positive for $u$ quarks, and negative for $d$ and $s$ quarks. These signs are consistent with the qualitative analysis in the BELLE experimental paper and Ref.~\cite{DAlesio:2020wjq}. 

Earlier extractions of the TMD PFFs are mainly from the $\Lambda$ polarization in proton-proton collisions~\cite{Anselmino:2000vs}, where a proper TMD factorization is not justified. It would be interesting to look into the connections between the data in these two different processes. We further make predictions for the transverse polarization of the $\Lambda$ and $\bar{\L}$ baryons produced in semi-inclusive deep inelastic scattering. The size of the polarization is around $10\%$, and thus should be measurable at the Electron Ion Collider. 

As we have mentioned, we look forward to future experimental data with higher statistics, and subsequently hope to disentangle the transverse-spin dependent pieces with differing azimuthal dependencies.

\section*{Acknowledgements}
We thank D.~Boer and H. Matevosyan for insightful communication, as well as K.~Lee and F.~Zhao for useful discussions. D.C. is supported by UCLA's University Research Fellows program, Z.K. is supported by the National Science Foundation under Grant No.~PHY-1720486, and J.T. is supported by the National Science Foundation under Grant No.~DGE-1650604.


\bibliography{refs}

\bibliographystyle{h-physrev5}

\end{document}